\documentclass[a4paper]{jpconf}
\usepackage{graphicx}
\usepackage{amsfonts, amsmath, amsthm, amssymb,hyperref}
\usepackage[font=small,labelfont=bf]{caption}
\usepackage{subcaption}
\usepackage{verbatim}

\begin{document}

\vspace*{-2.7cm}
\title{Viscous cosmological fluids and large-scale structure }
\author{B.G Mbewe$^1$,  R.R Mekuria$^{2}$, S. Sahlu$^{1,3}$, and A Abebe$^{1, 4}$}
\address{$^1$Centre for Space Research, North-West University, Potchefstroom 2520, South Africa}
\address{$^{2}$ Ala-Too International University, Faculty of Engineering and Computer Science, Bishkek, Kyrgzstan}
\address{$^{3}$ Entoto Observatory and Research Center, Space Science and Geospatial Institute, Addis Ababa Ethiopia}
\address{$^{4}$ National Institute for Theoretical and Computational Sciences (NITheCS), South Africa}

\ead{bonang.mbewe@gmail.com}

\begin{abstract}
In this paper, we study the viscous fluid cosmological model that when certain conditions are invoked mimics the $\Lambda$CDM model. The background equations governing the evolution of viscous interacting fluids in a multifluid system are derived. The Markov Chain Monte Carlo (MCMC) simulation is applied to constrain the best-fit cosmological parameters with Supernova Type 1a data. In addition, linear cosmological perturbations are investigated in a dust-matter-dominated frame using a $1+3$ covariant formalism approach. It is evident from the perturbation results obtained that the model predicts the disintegration of bound structures of large-scale structures in the late-time universe.
\end{abstract}

\section{Introduction}

Recent cosmological observations have indicated that the universe is undergoing an accelerated expansion. This accelerated expansion of the universe phenomena associated with dark energy has opened up a plethora of dark energy theories that have been proposed in order to address the behaviour that the current universe is displaying. $\Lambda$CDM amongst others, is one of the models proposed to explain the history of cosmic acceleration taking a simple and straightforward form of dark energy as a constant (also known as the cosmological constant $\Lambda$ and cold dark matter (CDM)) \cite{bamba2012dark,copeland2006dynamics,shi2012comprehensive}. Although $\Lambda$CDM has proved successful in most observational tests, as scientific models often do, it has faced challenges in explaining certain phenomena, and as such more models for dark energy are proposed as an alternative to $\Lambda$CDM.\\

\noindent In this paper, we scrutinize a model that, when certain conditions are invoked, mimics $\Lambda$CDM. For a multifluid universe with viscous effects taken into consideration \cite{fabris2006bulk} and the dark sector components coupled together \cite{bolotin2015cosmological,chimento2000enlarged}, allowing for exchange of energy from one to the other, and lastly, the equation of state for dark energy is assumed to be given in an inhomogeneous form \cite{shi2012comprehensive,astorga2019compact,garcia2018brane,garcia2019cosmic,hernandez2019cosmological}.
In addition, the model will be used to test/study the observational feasibility and implications for the large-scale structure formation of the universe. First, after the background equations governing the model are obtained, the cosmological background parameters are to be constrained using Supernovae Type 1a data by use of MCMC simulation. Thereafter, linear cosmological perturbations will be studied in the $1+3$ covariant formalism for a multifluid system \cite{abebe2012covariant,ellis2012relativistic,hough2021confronting,sahlu2019chaplygin}. Moreover, the density perturbation equations and solutions will be derived, and the cosmological implications of the density perturbations on large-scale structure formation will be discussed.\\

\section{Background cosmology of viscous interacting dark fluids (VIDF).}

\noindent In a homogeneous and isotropic universe, the background equations for a viscous interacting fluid in a flat geometry setting of FLRW metric \cite{brevik2015dark} read as:
\begin{equation}\label{Fluid eqtn}
	\begin{split}
			&\dot{\rho}_{r} +3H(\rho_{r} + p_{r}) = 0,\\
			&\dot{\rho}_{d} + 3H(\rho_{d} + p_{d}) = Q,\\
			&\dot{\rho}_{\Lambda} + 3H(\rho_{\Lambda} + p_{\Lambda}) = -Q,\\
			&\dot{H} = -\frac{1}{2}(\rho_{tot} + p_{tot}).
	\end{split}
\end{equation}

\noindent Here a dot denotes a derivative with respect to cosmic time. $\rho_{r}$, $\rho_{d}$, $\rho_{\Lambda}$ and $\rho_{tot}$ refer to the energy density of radiation, dust, vacuum and total fluid while $p_{r}=\rho_{r}/3$, $p_{d} = 0$, $p_{\Lambda}$ and $p_{tot}$ refer to isotropic pressure terms of vacuum and total fluid. $H$ is the Hubble parameter, and $Q$ is the interacting parameter. Natural standard units have been adopted, that is $8\pi G = c = 1$, and will be used throughout this manuscript. Taking the source term as taken by Brevik \cite{Brevik_2012}:
\begin{equation}\label{Coupling parameter}
	Q = \delta H\rho_{d},
\end{equation} 

\noindent where $\delta$ is a dimensionless constant. However, it can be noticed from the setup of conservation equations given by Eq. (\ref{Fluid eqtn}) that when $\delta >0$, energy exchange flows from dark energy to dark matter and conversely for $\delta < 0$, while when $\delta = 0$ the dark fluids are decoupled from each other \cite{vanderwesthuizen2023interacting}. Moreover, taking the equation of state for dark energy to be given by an inhomogeneous form as:
\begin{equation}\label{EoS DE}
	p_{\Lambda} = w_{\Lambda}\rho_{\Lambda} - \zeta, \quad \textrm{where} \quad \begin{cases}
	w_{\Lambda} = A_{0}\rho_{\Lambda}^{\alpha -1}-1,\\
	\zeta = \zeta_{0}\rho_{\Lambda 0}\bigg(\frac{\rho_{\Lambda}}{\rho_{\Lambda 0}}\bigg)^{m}.
	\end{cases}
\end{equation}

\noindent By assuming that the dimensionless constants $\alpha = m = 1$, to simplify the fluid equation for dark energy, the reduced equation of state for dark energy, now reads:
\begin{equation}\label{EoS DE renewed}
	p_{\Lambda} = (A_{0}-1-\zeta_{0})\rho_{\Lambda},
\end{equation}

\noindent with $A_{0}$ and $\zeta_{0}$ being dimensionless constants. The solutions that govern the evolution of each fluid species dynamics, in a multi-fluid universe are obtained as follows:
\begin{equation}\label{Energy density}
	\begin{split}
		&\Omega_{r} = \frac{\Omega_{r0}}{h^{2}}(1+z)^{4} \quad ; \quad \Omega_{d} = 				 		\frac{\Omega_{d0}}{h^{2}}(1+z)^{3-\delta} \quad ; \\
		&\Omega_{\Lambda} = \frac{1}{h^{2}}\bigg[(1-\Omega_{r0})(1+z)^{3(A_{0}-\zeta_{0})} \\
            & + \frac{\Omega_{d0}}{3(1+\zeta_{0}-A_{0}) -\delta}\bigg\{\delta(1+z)^{3-\delta} - 3(1+	 	 		\zeta_{0}-A_{0})(1+z)^{3(A_{0}-\zeta_{0})}\bigg\}\bigg].
	\end{split}
\end{equation}

\noindent Here $\Omega_{r0}$, $\Omega_{d0}$ refer to the fractional energy densities for radiation and dust evaluated at present cosmic time. Furthermore, $h = H/H_{0}$ denotes a normalised Hubble parameter.

\begin{figure}[h!]
	\centering
	\includegraphics[width = 1\textwidth, height=0.7\textwidth]{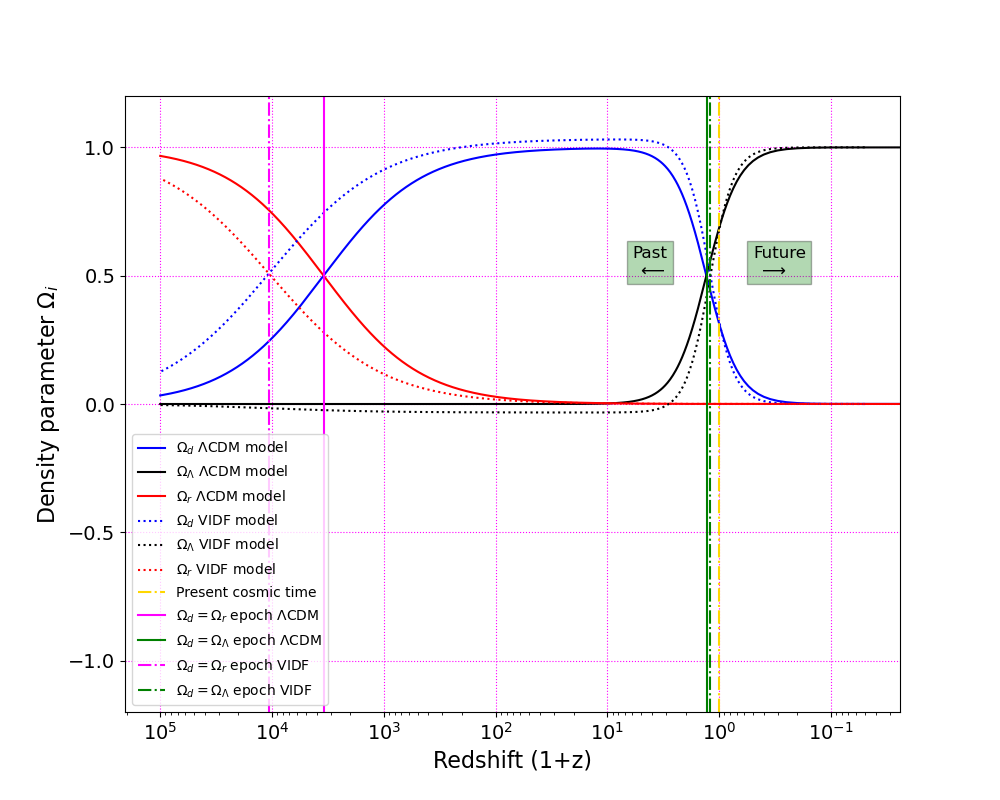}
	\caption{The evolution of energy density parameter for a multi-fluid universe.}
	\label{fig:Density_parameter}
\end{figure}

\noindent It is observed from Fig. \ref{fig:Density_parameter}, that the viscous interacting model compared to the $\Lambda$CDM model has the following features:
the radiation-dust matter equality occurs earlier, while the dust matter-dark energy equality epoch occurs later than $\Lambda$CDM. The dust-matter domination era is longer than that predicted by the $\Lambda$CDM model. Furthermore, the viscous interacting model violates all the energy conditions by having a dark energy density that is negative. The Hubble parameter governing the expansion of viscous dark-energy model is given as:

\begin{equation}\label{Hubble parameter}
	\begin{split}
	H = &H_{0}\bigg\{\Omega_{r0}\bigg[(1+z)^{4}-(1+z)^{3(A_{0}- 
            \zeta_{0})}\bigg]\\ 
            &+ \Omega_{d0}\bigg[\frac{3(1+\zeta_{0} - A_{0})}{3(1+\zeta_{0} - A_{0})-\delta}\bigg]	 		 	\bigg[(1+z)^{3-\delta}-(1+z)^{3(A_{0}-\zeta_{0})}\bigg] 
            +(1+z)^{3(A_{0}-\zeta_{0})}\bigg\}^{\frac{1}{2}}.
        \end{split}
\end{equation}

\begin{figure}[h!]
\centering
\begin{minipage}{0.5\textwidth}
  \centering
  \includegraphics[width=1\textwidth]{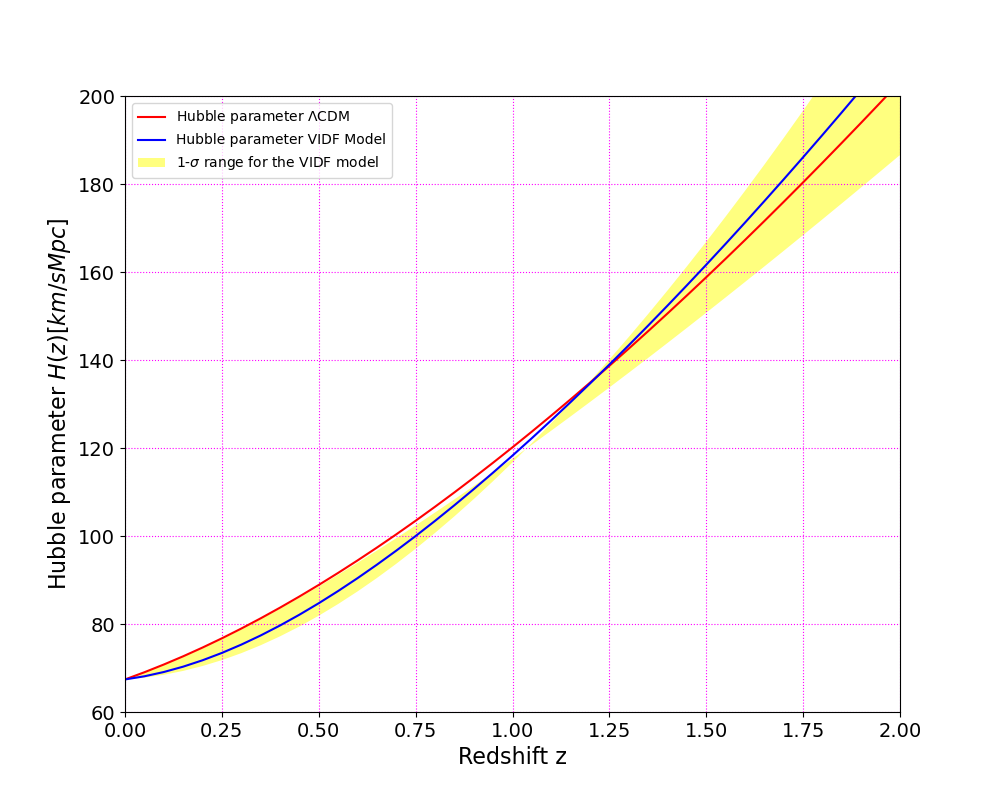}
  \captionof{figure}{The evolution of Hubble parameter}
  \label{fig:Hubble parameter}
\end{minipage}%
\begin{minipage}{0.5\textwidth}
  \centering
  \includegraphics[width=1\textwidth]{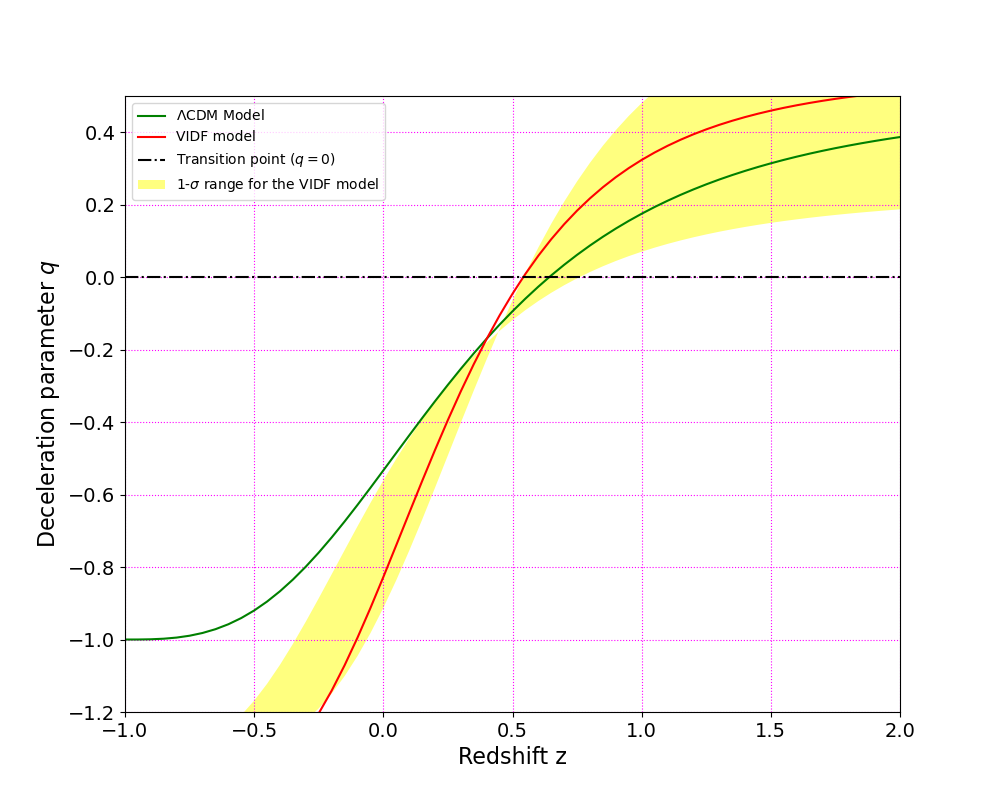}
  \captionof{figure}{The evolution of deceleration parameter}
  \label{fig:Decelration parameter}
\end{minipage}
\end{figure}

\noindent The deceleration parameter can be obtained  from Eq. \eqref{Hubble parameter} and it yields as:
\begin{equation}\label{Deceleration parameter}
    \begin{split}
            q(z) =& \frac{1}{2h^{2}}\bigg[\Omega_{r0}\bigg\{4(1+z)^{4} -               3(A_{0}- \zeta_{0}) 1+z)^{3(A_{0}-\zeta_{0})}\bigg\} \\
                  &+\Omega_{d0}\bigg(\frac{3(1+\zeta_{0} -A_{0})}{3(1+\zeta_{0}   -A_{0}) -\delta}\bigg)\bigg\{(3-\delta)(1+z)^{3-\delta} 
                  -3(A_{0} -\zeta_{0})(1+z)^{3(A_{0} -\zeta_{0})} \bigg\} \\
                  &+3(A_{0}-\zeta_{0})(1+z)^{3(A_{0}-\zeta_{0})}\bigg] - 1.
    \end{split}
\end{equation}
From Eq. \eqref{Hubble parameter}, the luminosity distance can be expressed as \cite{sahlu2019chaplygin,chanda2023observational}:
\begin{equation}
	\mu = 25+5\log_{10}\left[3000\bar{h}^{-1}(1+z)\int^{z}_{0}\frac{dz^{\prime}}{H(z^{\prime})/H_0}\right].
\label{eq: Distance modulus}
\end{equation}
\noindent In Fig. \ref{fig:Decelration parameter}, we see the transition from a decelerating universe to an accelerating universe occurring much later than that of the $\Lambda$CDM model, which was made evident in Fig. \ref{fig:Density_parameter} when the dust-dark energy equality epoch was later, unlike the predicted $\Lambda$CDM. Moreover, the two models seem to both predict the fact that the universe will indeed be in an accelerated regime indefinitely.

\section{MCMC results of the viscous interacting fluids}
\begin{figure}[h!]
\centering
\begin{minipage}{0.45\textwidth}
  \includegraphics[width=1\textwidth]{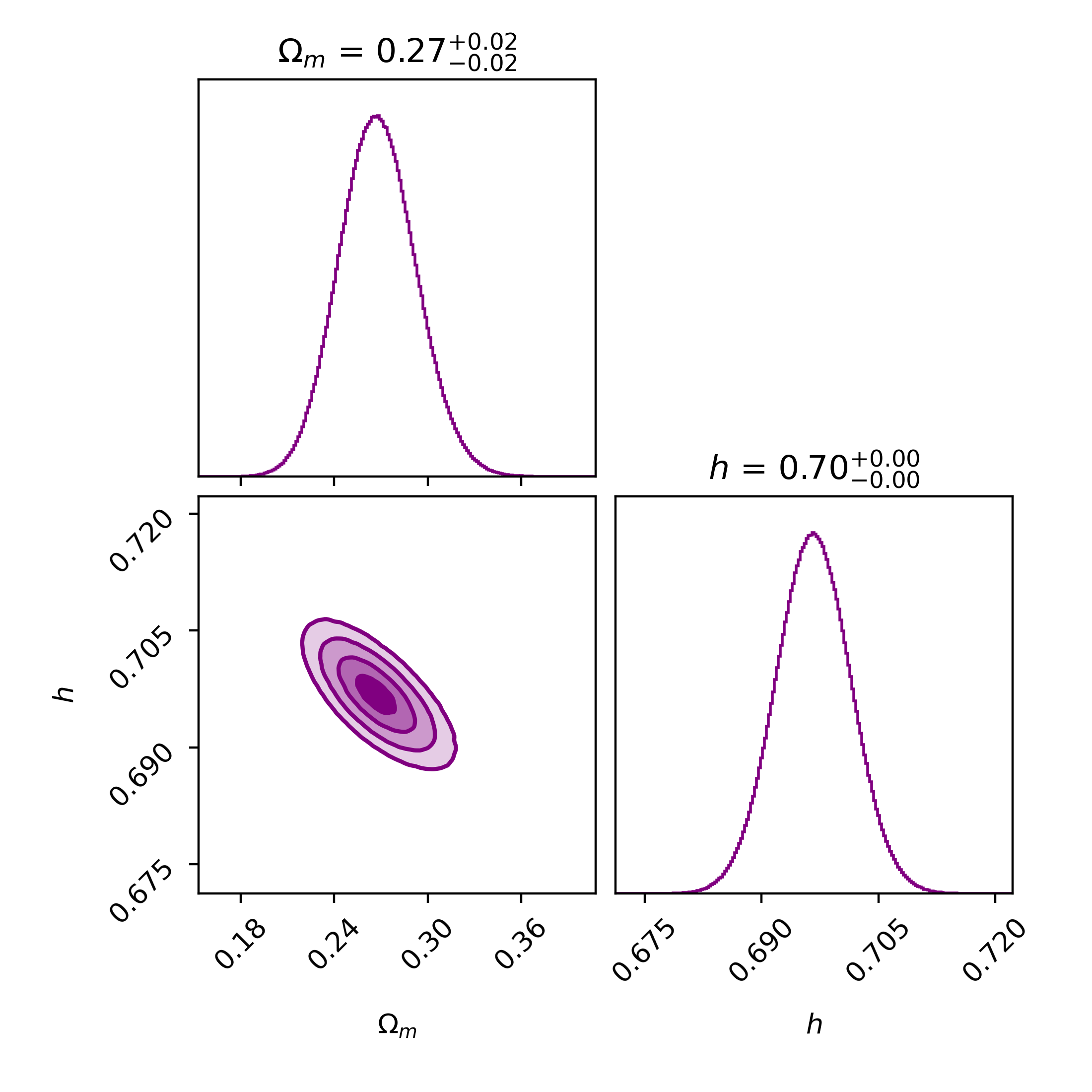}
  \captionof{figure}{Best fit parameters of $\Lambda$CDM model using MCMC.}
  \label{fig:LCDM MCMC}
\end{minipage}%
\qquad
\begin{minipage}{0.45\textwidth}
  \includegraphics[width=1\textwidth]{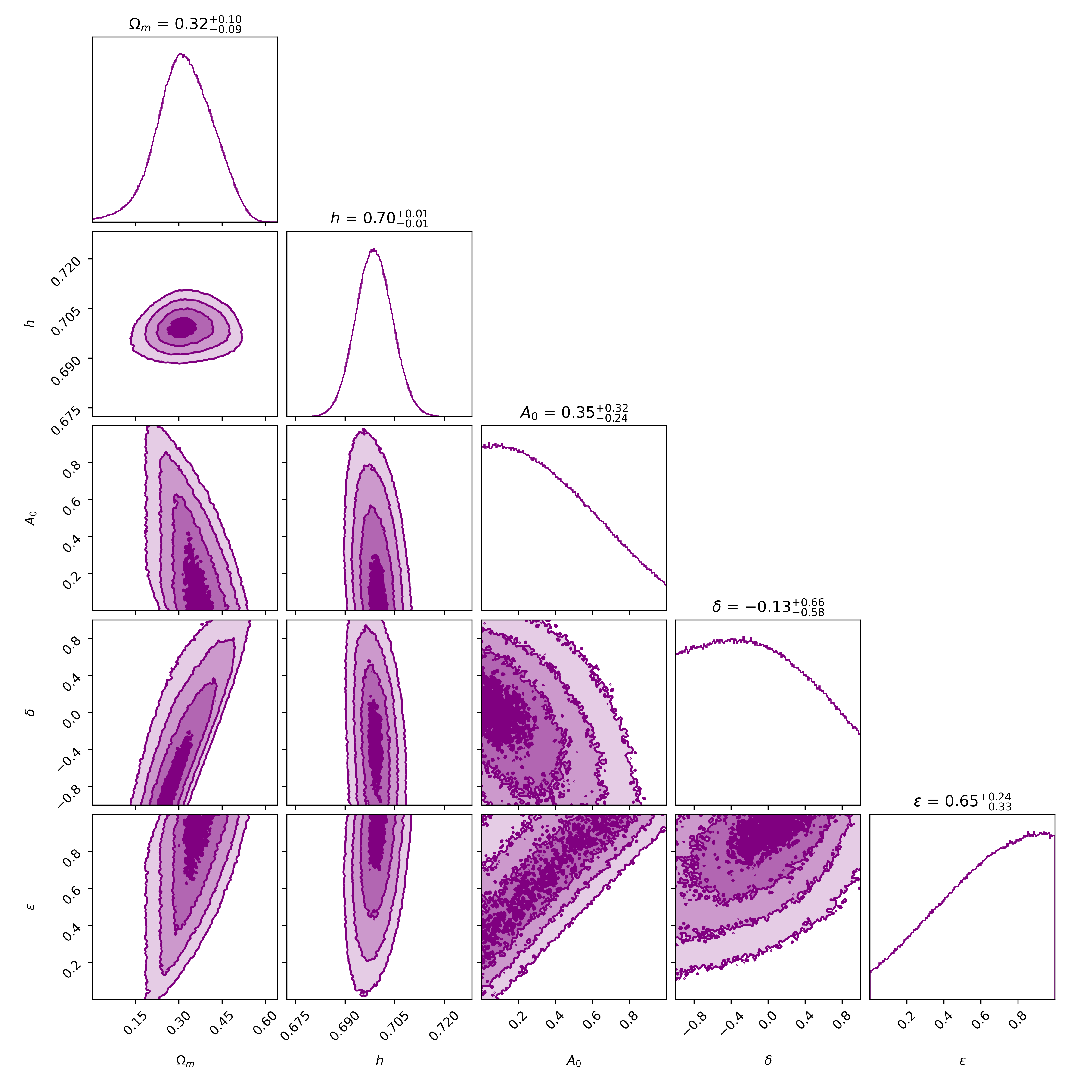}
  \captionof{figure}{Best fit parameters of VIDF model using MCMC.}
  \label{fig:VIDF MCMC}
\end{minipage}
\end{figure}

\noindent By applying the same methodology as detailed in \cite{hough2021confronting}, we constrain the best-fit cosmological parameter using Eq. \eqref{eq: Distance modulus}  with SNIa data, and the results are presented in Figs. \ref{fig:LCDM MCMC} and \ref{fig:VIDF MCMC} for $\Lambda$CDM and VIDF models, respectively.

\begin{comment}
\begin{figure}[h!]
     %\centering
     \begin{subfigure}[b]{0.5\textwidth}
         \centering
         \includegraphics[width=\textwidth]{LCDM_model_with SNe_results.png}
         \caption{Best fit parameters of $\Lambda$CDM model using MCMC.}
         %\label{fig:y equals x}
     \end{subfigure}
     \hfill
     \begin{subfigure}[b]{0.5\textwidth}
         \centering
         \includegraphics[width=\textwidth]{model_5_with SNe_results.png}
         \caption{Best fit parameters of VIDF model using MCMC.}
        % \label{fig:three sin x}
     \end{subfigure}
    \caption{The MCMC results of cosmological parameters.}
    \label{fig:two graphs}
\end{figure}
\end{comment}

\noindent In Fig. \ref{fig:VIDF MCMC} the constrained value of dust matter is in fairly good agreement with that obtained from Planck's 2018 data \cite{benetti2021dark}. However, as for the Hubble parameter, there exist some discrepancies as the value obtained is not within the $1\sigma$ deviation. In these MCMC simulation results, radiation was taken into account; however, it was treated as a constant, since when it is taken as a parameter to be constrained, the simulation fails to resolve it, and, as such, constraining it yields a uniform distribution over its entire search space, as outlined in detail by \cite{hough2021confronting}.

\section{Linear cosmological perturbations in VIDF model}

Taking the constrained cosmological parameters together with the background VIDF cosmology to study the growth of matter density fluctuations using a \begin{math}(1+3)\end{math} linear covariant cosmological perturbation to observe the contribution that the VIDF model has on large-scale structure formation. Using $(1+3)$ linear covariant formalism, we start by defining the spatial gradient variables \cite{ellis2012relativistic,sahlu2019chaplygin} as:
\begin{equation}
	D_{a}^{d} = \frac{a}{\rho_{d}}\nabla_{a}\rho_{d} \quad , \quad               D_{a}^{\Lambda} = \frac{a}{\rho_{\Lambda}}\nabla_{a}\rho_{\Lambda} \quad 
        , \quad Z_{a} = a\nabla_{a}\Theta,
\end{equation}

\noindent where $D_{a}^{i}$ is the inhomogeneity of the energy density of the species ($i = d,\Lambda$), $Z_{a}$ is the volume expansion of the fluids, and $\Theta = 3H$. Taking the cosmic time derivative, scalar decomposition and harmonic decomposition of the spatial gradient variables, we obtain the following:
\begin{equation}\label{Perturbation equations for VIDF}
    \begin{split}
        \Delta^{\prime}_{d} = \ & \bigg(\frac{3-\delta}{3(1+z)}\bigg)\frac{\mathcal{Z}}{h} + \bigg(\frac{3-\delta}{(1+z)}\bigg)\bigg(\frac{w_{d}\Omega_{d}}{(1+w_{t})\Omega_{t}}\bigg)\Delta_{d}
         +\bigg(\frac{3-\delta}{(1+z)}\bigg)\bigg(\frac{(w_{\Lambda}-\zeta_{0})\Omega_{\Lambda}}{(1+w_{t})\Omega_{t}}\bigg)\Delta_{\Lambda},\\
        \Delta^{\prime}_{\Lambda} = \ & -\Bigg[\frac{1}{3}\bigg(3(\zeta_{0}-A_{0})-\frac{\delta \Omega_{d}}{\Omega_{\Lambda}}\bigg)\Bigg]\frac{\mathcal{Z}}{(1+z)h}  \\
        \ & - \Bigg\{\Bigg[3(\zeta_{0}-A_{0})-\frac{\delta \Omega_{d}}{\Omega_{\Lambda}}\Bigg]\Bigg[\frac{w_{d}\Omega_{d}}{(1+w_{t})\Omega_{t}}\Bigg] -\frac{\delta \Omega_{d}}{\Omega_{\Lambda}}\Bigg\}\frac{1}{(1+z)}\Delta_{d}\\
        \ & -\Bigg\{\Bigg[3(\zeta_{0}-A_{0})-\frac{\delta \Omega_{d}}{\Omega_{\Lambda}}\Bigg]\Bigg[\frac{(w_{\Lambda}-\zeta_{0})\Omega_{\Lambda}}{(1+w_{t})\Omega_{t}}\Bigg] +\frac{\delta \Omega_{d}}{\Omega_{\Lambda}}\Bigg\}\frac{1}{(1+z)}\Delta_{\Lambda},\\
        Z^{\prime} = \ & \frac{2}{(1+z)}\mathcal{Z}\\
        \ & -\Bigg[\Bigg\{\frac{k^{2}(1+z)^{2}}{3h^{2}} -1 -\frac{1}{2}\bigg((1+3w_{d})\Omega_{d} +\big[1+3(w_{\Lambda}-\zeta_{0})\big]\Omega_{\Lambda}\bigg)\Bigg\}\frac{w_{d}\Omega_{d}}{(1+w_{t})\Omega_{t}}\\
        \ & \qquad-\frac{1}{2}(1+3w_{d})\Omega_{d}\Bigg]\bigg(\frac{3h}{(1+z)}\bigg)\Delta_{d} \\
        \ & -\Bigg[\Bigg\{\frac{k^{2}(1+z)^{2}}{3h^{2}} -1 -\frac{1}{2}\bigg((1+3w_{d})\Omega_{d} +\big[1+3(w_{\Lambda}-\zeta_{0})\big]\Omega_{\Lambda}\bigg)\Bigg\}\frac{(w_{\Lambda}-\zeta_{0})\Omega_{\Lambda}}{(1+w_{t})\Omega_{t}}\\
        \ & \qquad-\frac{1}{2}\big[1+3(w_{\Lambda}-\zeta_{0})\big]\Omega_{\Lambda}\Bigg]
        \bigg(\frac{3h}{(1+z)}\bigg)\Delta_{\Lambda}. 
    \end{split}
\end{equation}

\noindent Here, $\Delta_{m}$, $\Delta_{\Lambda}$ and $\mathcal{Z}$ are the scalar perturbations for dust matter, dark energy, and volume expansion. The prime denotes the derivative with respect to the redshift. It is seen from Eq. \eqref{Perturbation equations for VIDF} that conditions required for the $\Lambda$CDM model retrieval, that indeed even on the perturbation level is also the case \footnote{All the derivations of the equations found in this proceedings paper can be obtained in this Google Drive link \url{https://drive.google.com/drive/folders/1h9K_JW36rUv8Yti6befMkf4U-PKqWxyO}}.\\

\begin{figure}[h!]
     %\centering
     \begin{subfigure}[b]{0.3\textwidth}
         \centering
         \includegraphics[width=\textwidth]{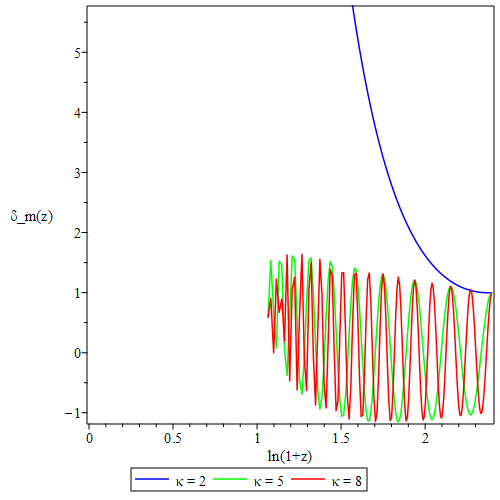}
         \caption{ Central values taken of MCMC.}
         \label{fig:central}
     \end{subfigure}
     \hfill
     \begin{subfigure}[b]{0.3\textwidth}
         \centering
         \includegraphics[width=\textwidth]{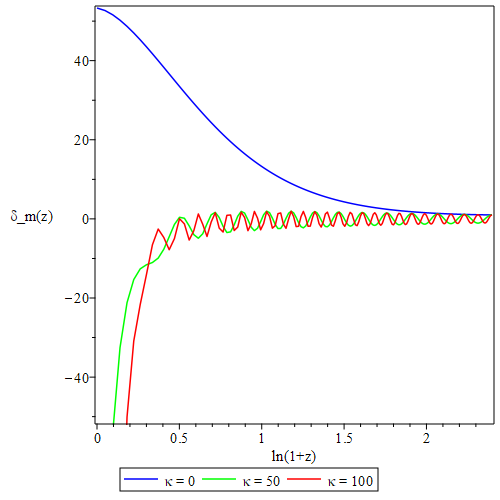}
         \caption{ Upper boundary values taken of MCMC.}
         \label{fig:upper}
     \end{subfigure}
     \hfill
     \begin{subfigure}[b]{0.3\textwidth}
         \centering
         \includegraphics[width=\textwidth]{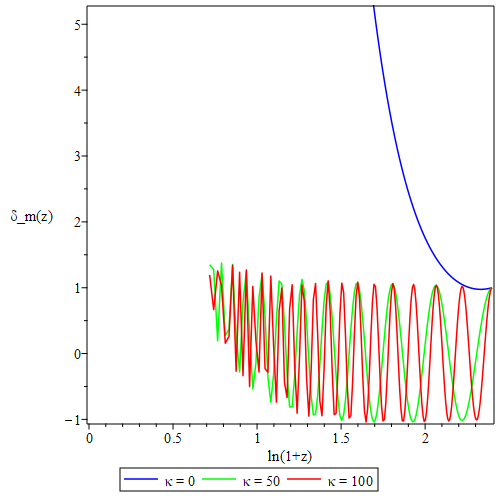}
         \caption{ Lower boundary values taken of MCMC.}
         \label{fig:lower}
     \end{subfigure}
        \caption{The density contrast of VIDE model as a function of redshift.}
        \label{fig:Perturbations}
\end{figure}

\noindent Fig. \ref{fig:Perturbations}, gives different extrema displayed by the VIDF model. In Fig. \ref{fig:central}, the central values obtained from MCMC simulation were used to obtain the density contrast of dust matter, and it can be seen for long wavelengths that the energy density fluctuations growths faster, however, there seems to be a singularity at a redshift of ($\sim 2$), while for short wavelengths the energy density seems to oscillate with growing energy density fluctuations, that is the same behaviour noticed when considering the lower boundary values obtained from the simulation. However, in Fig. \ref{fig:upper}, there exist no singularity which was experienced by the model in Figs. \ref{fig:central} and \ref{fig:lower} and we thus see the growth of matter density fluctuation throughout redshift for longer wavelength while a decaying one when considering short wavelengths.

\section{Conclusion}

\noindent In this paper the VIDF model has been explored in the FLRW universe where homogeneity and isotropy are implied and compared with the $\Lambda$CDM model. In the background cosmology of the VIDF model, it is noted that the era of dust matter-dominated epoch is much longer in comparison to the $\Lambda$CDM and also that the radiation-dust matter equality occurs much earlier, while dust-dark energy density equality occurs at a later cosmic time. Furthermore, the transition point from decelerated expansion to accelerated expansion occurs much later than the one anticipated by the standard model. Both models seem to agree when it comes to accelerating expansion of the universe prediction.\\

\noindent MCMC simulation was adopted in constraining the model's background cosmological parameters using supernova data. The results obtained by constraining the cosmological parameters of the model were more in agreement with the recent cosmological data given by Planck's 2018 data. Although, the were more models studied, we did not consider them for the proceeding since some of them were not giving results at all while others did not give comparable results to that of Plank's 2018 by overestimating. However, the results obtained are preliminary, as more data (such as BAO, CMB, OHD, R22 ect.) have to be tested on the model in order to see how the model fits the observation. The linear cosmological perturbations were also investigated in a dust matter-dominated frame using a $1+3$ covariant approach. It was evident from the perturbation results obtained that the model predicts the rip of large-scale structures at the late-time universe, and this comes mainly from the interaction of dark-fluid components. Future work will be to use different cosmological data to constrain the background parameters and large-scale structure even better.\\
\newpage
\section*{References}

%\bibliographystyle{iopart-num}
%\bibliography{iopart-num}

%Start the appendix [OPTIONAL]

\providecommand{\newblock}{}

\end{document}